%Paper: hep-th/9401073
%From: Michael Douglas <mrd@mike.rutgers.edu>
%Date: Fri, 14 Jan 94 15:52:55 -0500

\input harvmac.tex
\def\Title#1#2{\rightline{#1}
\ifx\answ\bigans\nopagenumbers\pageno0\vskip1in%
\baselineskip 14pt plus 1pt minus 1pt
\else%\special{papersize=11in,8.5in}%
\def\listrefs{\footatend\vskip 1in\immediate\closeout\rfile\writestoppt
\baselineskip=14pt\centerline{{\bf References}}\bigskip{\frenchspacing%
\parindent=20pt\escapechar=` \input refs.tmp\vfill\eject}\nonfrenchspacing}
\pageno1\vskip.8in\fi \centerline{\titlefont #2}\vskip .5in}

% *************************************

\def\ul#1{{\it #1}}
\def\vev#1{\langle{#1}\rangle}

\def\tr{{\rm tr}}
\def\cr{\hfill\break}
\vskip .65truein
\noindent
\Title{\vbox{\baselineskip12pt\hbox{RU-94-9}}}%
{\vbox{\centerline{Some Comments on QCD String}}}
\centerline{Michael R. Douglas}
\centerline{Dept. of Physics and Astronomy}
\centerline{Rutgers University}
\bigskip
\bigskip
\bigskip
\noindent
We try to draw lessons for higher dimensions from the string representations
recently derived for large $N$ Yang-Mills theory by Gross and
Taylor, Kostov, and others,
and call attention to three characteristics that should be expected of a string
theory precisely equivalent to a higher dimensional gauge theory:
continuous world-sheets; strong coupling at short distances; and negative
weights.\cr
To appear in the proceedings of the Strings '93 Berkeley conference.
\Date{January 1994}
\nref\Gross{D. J. Gross, Nucl. Phys. B400 (1993) 161.}
\nref\DK{M. R. Douglas and V. A. Kazakov, to appear in Phys. Lett. B.
hep-th/9305047.}
\nref\GT{D. J. Gross and W. Taylor, Nucl. Phys. B400 (1993) 181 and
 Nucl. Phys. B403 (1993) 395.}
\nref\Kostov{I. K. Kostov, SACLAY-SPHT-93-050, June 1993. hep-th/9306110.}
\nref\MR{A.A.Migdal, Sov. Phys. JETP 42 (1975) 413, 743.\cr
B.Rusakov, Mod.Phys.Lett. A5 (1990) 693.}
\nref\Rusakov{B.Rusakov, Phys.Lett. B303 (1993) 95.}
\nref\cargese{V. A. Kazakov, M. R. Douglas, to appear in the proceedings of
the May 1993 Carg\`ese workshop on Strings, Conformal Models and Topological
Field Theories.}
\nref\others{J. A. Minahan and A. P. Polychronakos,
CERN-TH-7016-93, Sept. 1993.  hep-th/9309119. \cr
M. Caselle, A. D'Adda, L. Magnea, S. Panzeri, DFTT-50-93,
Sept. 1993. hep-th/9309107.\cr
J.-M. Daul and V. A. Kazakov, LPTENS-93-37, Oct. 1993.
hep-th/9310165.\cr
D. V. Boulatov, NBI-HE-93-57, Oct. 1993.  hep-th/9310041.}
\nref\kkone{V. A. Kazakov and I. K. Kostov, Nucl. Phys. B176
(1980) 199.}
\nref\Polchinski{J. Polchinski, UT Austin preprint UTTG-16-92,
hep-th/9210045, and references there.}
\nref\Weingarten{D. Weingarten, Phys. Lett. 90B (1980) 280.}
\nref\minlength{I. Klebanov and L. Susskind, Nucl. Phys. B309
(1988) 175;\cr
K. Konishi, G. Paffuti, and P. Provero, Phys. Lett. B234 (1990) 276.}
\nref\Polyakov{A. M. Polyakov, Nucl.Phys. 268B (1986) 406;\cr
A. M. Polyakov, ``Gauge Fields and Strings,'' Harwood (1987).}
\nref\NielsBohr{
B. Durhuus, J. Fr\"ohlich and T. Jonsson, Nucl. Phys. B240[FS12] (1984) 453;\cr
J. Ambjorn and B. Durhuus, Phys. Lett. 188B (1987) 253;\cr
J. Ambjorn, in Random Surfaces and Quantum Gravity, Plenum 1991, pp. 327-336.}
\nref\AIJP{J. Ambjorn, A. Irback, J. Jurkiewicz,
B. Petersson Nucl. Phys. B393 (1993) 571-600.}
\nref\membrane{``Statistical Mechanics of Membranes and Surfaces,''
vol. 5, eds. D. Nelson, T. Piran and S. Weinberg, World Scientific (1989);\cr
F. David, B. Duplantier, S. Leibler, L. Peliti, Phys. Rep. 184 pp. 221-282.}
%\nref\GW{D. J. Gross and E. Witten, Phys. Rev. D21 (1980) 446.}
\nref\Douglas{M. R. Douglas, RU-93-57, Oct. 1993.
hep-th/9311130.}
\nref\HighT{J. Polchinski, Phys. Rev. Lett. 68 (1992) 1267.}
Other speakers at this conference have reviewed the long history of attempts to
precisely reformulate large $N$ $SU(N)$ gauge theory as a string theory, and
the recent line of attack, initiated by D. Gross,\Gross\ who has proposed
that since we expect QCD string, if it exists, to exist in dimensions
$2\le D\le 4$, and since Yang-Mills theory in $D=2$ is exactly solvable, it
behooves us to understand that case as fully as possible, by reformulating it
as a string theory and attempting to identify features which generalize to
arbitrary $D$.

My own recent work, with V. Kazakov,\DK\ has concentrated on understanding
the validity of the string representation of \GT\ in $D=2$ for the case
which seems to us most representative of higher dimensions, namely where
two-dimensional (Euclidean) space-time is a sphere.
We consider this case more representative than the other Riemann surfaces
studied in \GT\ because it is the only case in which the free energy
displays the $O(N^2)$ behavior we expect in higher dimensions, and the only
case in which a saddle point dominates the functional integral, recovering the
usual reason for expecting the large $N$ limit to be simple.

The basic result is very simple.
One has the formula of Migdal and Rusakov\MR\ for the partition function
on a sphere with area $A$ as a sum over representations, which labelling
representations by the distinct integers $n_i=\alpha_i+\rho_i$ (the highest
weight plus half the sum of positive roots in a diagonal Cartan basis) becomes
\eqn\parti{
  \lim_{N\rightarrow\infty}
Z_{G}(A) = \exp[N^2 F(A)] =
  \sum_{n_i\ne n_j} \prod_{i>j} \left({n_i - n_j \over i-j } \right)^{2-2G}
  \exp{- {g^2 A \over 2 N} \sum_{i=1}^{N} n_i^2}.}
Since the $n_i$ are distinct they are $O(N)$ and the sum
$(1/N)\sum n_i^2 \sim N^2$.
For the case of $G=0$ only we can regard the prefactor as an `entropic' term of
$O(\exp N^2)$ which competes with the exponential to make the total summand
highly peaked for a single `master representation' with particular values
$n_i$.
Minimizing the resulting effective action is the same problem as for an
integral over an $N\times N$ hermitian matrix, with a single difference:
in both cases we can describe the saddle point distribution by a continuous
density $\rho(n/N)$, but in the present case the discreteness of the variables
$n_i$ implies a maximum bound on the density,
\eqn\bound{\rho(n/N) \le 1.}
Disregarding this constraint, the saddle point is the standard one for a
Gaussian integral:
\eqn\wi{\rho(x) = {g^2 A \over 2 \pi } \sqrt{ {4 \over g^2 A} - x^2 }. }
However, for $g^2 A\ge g^2 A_{crit} = \pi^2$, this saddle point violates the
constraint.  It is not hard to enforce the constraint by hand and find another
saddle point, with density expressed in terms of elliptic functions.\DK

The string representation of \GT\ is a valid series expansion for this
second, strong coupling phase.
The weak coupling answer (first found this way in \Rusakov)
is very much simpler and not immediately suggestive of any string
representation.

Much more can be said about this problem,%
\footnote{$^*$}{And most of my lecture at Berkeley was devoted to this,
but that material will be published in \cargese.
See as well \others.
The remainder of this article is based on my introductory comments at
Berkeley, expanded and presented in talks at ENS, CERN and Princeton in July
1993.}
but I will just mention the main lesson I drew from this work
(which I admit may be too pessimistic):
it is that the evidence is against any precise correspondance between the
present string constructions (in any $D$) and weak coupling, continuum
Yang-Mills theory.

On the other hand, one feels that these string constructions do capture
something real about gauge theory, and as such deserve interpretation, in
particular hypotheses about which features survive the continuum limit
(both world-sheet and space-time)
and are responsible for the drastic differences between QCD string and the
known critical and non-critical strings.
In fact, after trying to make sense of the idea of QCD string,
the known critical and non-critical strings start to all look so similar that I
will lump them all together in the following under the rubric `fundamental
string.'
(What I mean by this will become more clear below.)
My own thinking has largely been in the context of \refs{\GT,\Kostov,\kkone}
and in particular about features which are apparent in any $D$ in
the formalism of \Kostov~
but I suspect that other derivations of string representations would also
suggest these hypotheses, especially the first one.
I also believe that what I have to say is already known to some workers in the
field, but since it has not been written down in one place, it seems of value
to collect it here.
Let me also mention that many interesting
statements have been made about QCD string which I will not repeat here; see
\Polchinski\ for a review of some of these.

On the most basic level are statements which are not specific to Yang-Mills
theory but would also be true for `Weingarten theory' (a lattice theory with
link variables allowed to take arbitrary complex hermitian values and a
quadratic term in the action to make this integral well defined).%
\Weingarten~
A striking feature of the string representations (and of the strong coupling
expansion at finite $N$) is that the partition function is expressed as a sum
over \ul{continuous} embeddings of surfaces in space-time.
Now continuity is certainly part of the definition of the word `embedding'
and one might think it only natural for a string theory.
But, as is well known,\minlength\ it is not true for the fundamental
string, here taken to be any string with embedding described by world-sheet
variables $X^\mu(z)$ and an action whose short-distance behavior is controlled
by the term $(1/\alpha')\int d^2z~(\partial X)^2$.

What do we mean by this?
A continuous embedding $X^\mu(z)$
by definition satisfies the following:
for any $\epsilon>0$, there exists a $\delta$ such that
$|z-z'|<\delta$ implies $|X(z)-X(z')|<\epsilon$.
If we are integrating over embeddings, we can take this statement over to the
expectation values $\vev{|X(z)-X(z')|}$ or $\vev{(X(z)-X(z'))^2}$
if we have a uniform bound $\delta\ge\delta_0>0$ for all embeddings.

For the fundamental string, we can crudely estimate
\eqn\estimate{
\vev{(X(z)-X(0))^2} \sim z^2 \vev{\partial X(z)~\partial X(0)} \sim \alpha'}
which does not go to zero as $z\rightarrow 0$.
This crude argument can be improved in many ways
(it is a well-known fact of quantum field theory that at and above the critical
dimension $2$ continuous configurations do not dominate the functional
integral)
and in the context of strings for quantum gravity this fact has been
offered as a sense in which string theory has a `minimum length'
$\sqrt{\alpha'}$ below which conventional ideas of space-time are not
appropriate.\minlength

We see that we should not get such a result by taking a limit of a sum over
continuous embeddings, as we expect for QCD string.
To make this precise we would need to establish the uniform bound referred to
above, and to do this we would need a natural choice of local coordinate $z$ on
the world-sheet.
This brings us to a related point: it seems that the induced metric
$g_{ab} = \partial_a X^\mu \partial_b X_\mu$ plays a fundamental role in QCD
string.
This is clearest in the string representation of \Kostov\ (or any
string modelled on the lattice strong coupling expansion).
We can think of the lattice as embedded in $D$-dimensional flat space and
literally induce the metric and local coordinates onto each plaquette of a
surface.  Such a string manifestly satisfies the bound $|X(z)-X(z')| \le
d(z,z')$ where $d$ is the minimum world-sheet distance between two points.
More generally, we should expect the coordinate $z$ to be compatible with the
induced metric in the very minimal sense that we can locally write $g_{ab} ds^a
ds^b = e^\phi dz d\bar z$ with continuous $\phi$,
in which case we have a bound of the sort described.

How can we replace the very natural world-sheet action $(1/\alpha')\int
d^2z~(\partial X)^2$ ?
It is not hard to come up with candidates which satisfy continuity as defined
above.
A very simple one is just $\int \sqrt{h} (\Delta X)^2$
where $\Delta$ is the two-dimensional Laplacian for the metric $h$.
The metric $h_{ab}$ may be related to $g_{ab}$, but to make the point at hand
let's make it independent and non-singular around a point $z=0$, so we can use
a coordinate in which $h_{ab} ds^a ds^b \sim dz d\bar z$,
and the world-sheet theory is still free.
(Perhaps this can be thought of as the zeroth order in an expansion around
$h_{ab}=\vev{g_{ab}}$.)
Then the short distance behavior we expect is
\eqn\less{\vev{X(z) X(0)} \sim z^2 \log z}
which is compatible with continuity.

Another well-known difference between fundamental string theory and field
theory (and therefore with QCD string) is that in the former we cannot easily
define correlation functions local in space-time.  The simplest argument for
this is that a vertex operator must be dimension $2$ so that its world-sheet
integral will be covariant, and that the anomalous dimension of a vertex
operator $O(z,\bar z):e^{ikX(z,\bar z)}:$ has a contribution $\alpha' k^2$
which is fixed by this constraint.
We find it significant that the above considerations of continuity drastically
change this constraint as well.
If the Green's function $\vev{X(z) X(0)}$ is non-singular at short distance,
the operator $e^{ikX(z,\bar z)}$ does not need renormalization and therefore
has no anomalous dimension.
Thus we can define integrated world-sheet correlation functions for any
external momenta, and Fourier transform them in the usual way to get local
correlation functions in space-time.

Although I appealed to the strong coupling expansion,
I believe that the property of continuity has a more basic origin,
and is already strongly suggested by the simple observation that the
gauge-invariant observables are Wilson loops, which must be continuously
embedded in space-time.

The second hypothesis I would propose is much more closely tied to the strong
coupling expansion, and furthermore assumes that the QCD string does reproduce
the correct short-distance physics (e.g. asymptotic freedom in $D=4$).
It is that the QCD string, as a two-dimensional quantum field theory on the
world-sheet, is strongly coupled at short distances.
This would certainly explain why nobody has yet found the appropriate string
theory!
More generally, weak bare coupling in field theory, implies strong coupling on
the world-sheet for the string theory.
The observation behind this is the following:
we know that short distance QCD physics is well described by perturbation
theory, and that the link variables fluctuate but only very near the identity
(up to pure gauge fluctuations).
To get such a result out of the strong coupling expansion, we need to first
reproduce the zeroth order behavior, in other words the rough form
$\exp -\tr U/g^2$ of the action in this limit.
Now if the strong coupling expansion converged this would certainly work, but
we would clearly need a growing number of terms in the expansion as
$g\rightarrow 0$, and no one term in the expansion would dominate.
But, reinterpreting the expansion as a sum over world-sheets, this is just the
situation we would interpret as strong coupling.  In a weakly coupled theory a
semiclassical treatment of the path integral would be valid, meaning that a
single configuration (or a finite-dimensional space of configurations) would
dominate, and we would expand in fluctuations around this.
This is not the situation here.

Another way to see this is to observe the close similarity between the
derivation of the string (or strong coupling) expansions, which start from a
character expansion of the Boltzmann weights,
and the duality transform in abelian gauge theory.
In abelian gauge theory, once we do the link integrals, we have precisely
reformulated a weakly coupled theory of link variables as a strongly coupled
theory of the dual `representation' variables.
In non-abelian gauge theory, we have additional information to keep at this
point, which in the string representation is what defines the connectivity of
the world-sheet, but the strong-weak coupling relation is still present.

The two-dimensional case might at first be thought to contradict this, if we
accept the popular hypothesis that a continuum world-sheet theory exists
for which the functional integral can be reduced to a sum over classical
solutions, reproducing the explicit sum of \parti.
However this in itself is not incompatible with strong coupling
(as evidenced by a topological field theory with partition function independent
of $\hbar$)
and we might ask whether the $g^2 A\rightarrow 0$ limit is such that no single
classical solution dominates.  A good example where this is true is the
$O(N^0)$ free energy for the torus space-time,
with asymptotic behavior\refs{\Gross,\Douglas}
\eqn\torus{
F_{G=1} = - 2\log \eta(ig^2 A) \sim \log g^2 A + \pi/6g^2 A.}
Even the sphere is an example, though the large $N$ phase transition means that
the limit of the string answer need not reproduce the `correct' weak coupling
asymptotics (it behaves as $F \sim 1/(g^2 A)^2$, and is real, surprisingly
enough.)

The reader may have observed that my second hypothesis tends to undercut the
example (the free, higher derivative action) I gave to illustrate the first
hypothesis.  I would agree and consider this a failing of the example, not the
hypothesis.

Finally, I want to recall the results of the work on sums over discretized
surfaces with positive weights, especially the work of \NielsBohr.
There are very general arguments that in $D\ge 2$ such strings will always have
a `branched polymer' critical behavior.  Here is a clear difference between a
string derived from the Weingarten model and a QCD string, which cannot have
branched polymer behavior.
Now the easiest way to escape the arguments of \NielsBohr\ (as they
point out) is simply to give negative weights to some surfaces, and the string
representations of \refs{\GT,\Kostov} have negative weights in profusion.
In the $D=2$ theory of \GT\ the solution to the branched polymer
problem is not this, however.
Rather, it is that they explicitly restrict their sum over surfaces to
embeddings of the world-sheet in the target space without folds.

In what sense can this restriction generalize to arbitrary $D$ ?
If it does, could it solve the branched polymer problem in $D>2$ ?
If one simply thinks of excluding configurations which fold from the
statistical sum, since the condition can be stated locally on the world-sheet,
general considerations suggest that in the continuum limit this constraint
would renormalize to a local operator expressible in terms of the embedding
variables $X^\mu$.  Much attention has been devoted to the search for new
theories with embedding (and no other) degrees of freedom, especially in the
condensed matter (membrane) context.%
\membrane

A possible clue is given by the formalism of \Kostov.
There the `no-fold' constraint is not explicit but emerges after cancellations
between surfaces of positive and negative weight.
This formalism applies in $D>2$, and there it is not known how to state the
result as a simple cancellation: rather, it appears that some surfaces must
enter the sum with negative weights.

Another indication of the role of negative weights follows from an important
feature of \refs{\GT,\Kostov} not yet mentioned:
the world-sheet cosmological constant can be identified with the bare gauge
theory coupling constant.
This is one of the most significant differences between this expansion and the
strong coupling expansion derived from the Wilson action.
For readers more familiar with the $D=2$ case, the starting point for the $D>2$
case is to write each plaquette Boltzmann weight as a sum over terms each with
an interpretation as a cover by some number of string world-sheets with
specified topology.  The parameter $g^2 a^2$ for a plaquette of area $a^2$ in
two dimensions is replaced by $g_0^2 a^{4-D}$ where $g_0$ is the bare coupling
and $a$ the lattice spacing.
In $D=2$ the reproducing property of the heat kernel action implies that this
description is equally valid on all scales including the total system, and the
partition function will be a sum of terms corresponding to an $n$-fold cover to
the total space with weight $\exp -g^2 n A$ (times polynomial corrections).
In $D>2$ this is not true but string world-sheets are still formed by sewing
the covers of each plaquette, and a world-sheet covering plaquettes of total
area $n a^2$ (with multiplicity, i.e. a locally $m$-fold cover counts $m a^2$)
has weight $\exp -g_0^2 a^{4-D} n$ times polynomial corrections.
If we assume that the exponential dependence is the important dependence,
clearly we want to call $n$ the bare world-sheet area and $g_0^2 a^{4-D}$
the `bare cosmological constant'.
The simplest estimate for the partition function at fixed bare world-sheet area
would then be to count each cover once, giving the asymptotic behavior $\exp c
n$ for some positive $c>0$.
The constant is determined by the choice of lattice and
local considerations, for example whether we exclude folds.
(The exponential growth would have been present even in counting random walks,
and in $D>2$ any reasonable class of surfaces will include `fat random walks'
where a cutoff-scale loop traces a random walk.)

If we assume that a string representation of this type converges to correct
weak coupling answers, we can work backward from our weak coupling expectations
(in particular those given by the RG) to determine the dependence of the string
partition function on bare world-sheet area.  The relation is just Laplace
transform, and a simple example to make the point is to imagine that there is a
scaling term in the $D=4$ (with total volume $L^4$) free energy such as $F \sim
(L/a)^4 \exp -\beta/g_0^2$.  This could be the Laplace transform of a term in
the fixed area partition function such as $Z(A)\sim\cos \sqrt{A}$.  (The
subexponential dependence on $A$ is more significant than the non-positive
definite nature of this result; although both features are necessary to get
this $F(g_0^2)$, there could also be additive non-universal contributions to
$F$ which make it positive.)

The essential point is that with positive weight for each configuration,
the usual exponential growth in the number of configurations as a function of
the bare area will lead to a critical point at finite bare coupling.
As in \NielsBohr, this implies that the string tension remains at
the cutoff scale, preventing a continuum string interpretation (and generically
leading to branched polymer behavior).  It is clearly not the QCD critical
point at zero bare coupling.  A theory with the critical behavior appropriate
to QCD must have negative Boltzmann weights to produce a subexponential fixed
area partition function.  The no-fold constraint in higher dimensions does not
suffice.

\smallskip
To summarize, there are major differences between QCD string (if it exists) and
all known consistent string theories.
The most striking differences follow if we try to reproduce field-theoretic
short distance behavior.
One way out is to look for a string which only reproduces long-distance
physics, and whose short-distance physics is different from field theory.
If one thinks of the large $N$ phase transition on the two-sphere as a
non-analyticity of an observable as a function of length scale,
it might be taken as evidence that such a string exists, reproducing the
analytic continuation of observables (such as the Wilson loop expectation value
as a function of enclosed area) from their long distance behavior.
This is to be contrasted to the more traditional interpretation of the
transition as associated with a critical coupling and therefore posing a
barrier to any contact with the weak coupling continuum limit.
It would be important to find a tractable higher dimensional calculation to
distinguish between these scenarios.
A good example would be the finite temperature free energy;
the hope would be that the higher dimensional analog of the two dimensional
large $N$ transition was the deconfinement transition, and that a string
representation could reproduce the high temperature continuation of the
confining phase \HighT\ even at weak coupling.

On the other hand it might really be that a string theory exists with very
different short-distance properties than the fundamental string.
An example of such a theory may be the `rigid' string with an extrinsic
curvature term in the action, as advocated in \Polyakov, and we argued
that theories of this form are more likely than fundamental string theory to
reproduce certain qualitative features of field theory, such as the existence
of Green's functions of local operators.
To our present understanding, this string suffers a fatal flaw:
it is non-unitary.  Related to this, the classical theory is unstable.
We have no new answer to these problems, but we wonder if their ultimate
resolution might be as subtle as the problem of showing that non-abelian gauge
theory was renormalizable and unitary once was.

The two-dimensional examples allow us to make precise statements but do not
capture all the physics, even qualitatively.
Perhaps it would be fruitful to consider a larger class of models,
for example gauge theory defined on lattices more general than discretizations
of $D$-dimensional space, in hopes of finding tractable models illustrating the
points discussed above.

\medskip
I am indebted to many people for helping me learn about this subject,
and I would like to express my special thanks to I. Kostov for explaining his
work, and to V. Kazakov and M. Staudacher for their unflagging interest.
I would also like to express my thanks for the hospitality of the Institute for
Theoretical Physics, Santa Barbara, and of the Ecole Normale Sup\'erieure,
Paris.

\listrefs

\end